\documentclass[prl,aps,twocolumn,amsmath,amssymb]{revtex4-1}

\usepackage{multirow}
\usepackage{amssymb}
\usepackage{amsmath}
\usepackage{mathrsfs}
\usepackage{amsfonts}
\usepackage{color}
\usepackage{bm}
\usepackage{xspace,soul}

\usepackage{graphicx,subfigure}

\newcommand{\pref}[1]{(\ref{#1})}
\newcommand{\eref}[1]{Eq.~\pref{#1}}
\newcommand{\fref}[1]{Fig.~\pref{#1}}

\usepackage[absolute]{textpos}
\usepackage{calc}

\begin{document}

\title
{
Traces of integrability in scattering of one-dimensional dimers on a barrier 
}

\author{Juan Polo Gomez}
\email{juan.polo@lpmmc.cnrs.fr}
\affiliation{Univ.~Grenoble-Alpes, CNRS, LPMMC, 38000 Grenoble, France}

\author{Anna Minguzzi}
\affiliation{Univ.~Grenoble-Alpes, CNRS, LPMMC, 38000 Grenoble, France}

\author{Maxim Olshanii} 
\affiliation{Department of Physics, University of Massachusetts Boston, Boston, MA 02125, USA}

\begin{abstract}
%
We consider  molecules made of two one-dimensional short-range-interacting bosonic atoms. We show that in the process of scattering 
of these molecules off a narrow barrier, odd incident waves produce \emph{no unbound atoms, even when the incident energy exceeds the dissociation threshold}. 
This effect is a consequence of a prohibition on chemical reactions acting in a generally unphysical 
Bethe Ansatz integrable system of a $C_{2}$-type, with which our system shares the spatially odd eigenstates. We suggest several experimental implementations 
of the effect. We also propose to use  the monomer production as an alternative read-out channel in an atom interferometer: unlike in the standard interferometric schemes, 
no spatial separation of the output channels will be required. 

$\left.\right.$\\

\end{abstract}
%

\maketitle


%
The last two decades are marked by a revival of interest in Bethe Ansatz integrable particle systems
\cite{girardeau1960_516,lieb1963,mcguire1964_622,gaudin1983_book_english,korepin1997,sutherland2004_book,perelomov1990_book}, 
inspired by the 
emerged experimental relevance of the former in both many-body 
\cite{kinoshita2004_1125,paredes2004_277,tolra2004_190401,liao2010_567} and few-body \cite{moritz2005,wenz2013_6157} cold-atom systems (see 
\cite{batchelor2016_173001} for a review). Integrability, besides providing a way to produce theoretical predictions, also induces new empirically observable integrals of motion. Traditionally, the conservation of the momentum distribution is emphasised \cite{rigol2007,kinoshita2004_1125}. 
However, another experimentally sound set of conserved quantities is derived from the conservation of the chemical composition \cite{yurovsky2006_163201}, i.e.\
the decomposition of the system onto unbound atoms, dimers \cite{moritz2005}, trimers, all the way to the long polimers, the latter manifesting themselves as bosonic solitons \cite{strecker2002_150,khaykovich2002_1290}.

Below we show that an integrability-related chemical stability also appears in a system that consists of two one-dimensional attractive short-range-interacting bosons and a narrow barrier.   
While generally not integrable, our system shares some of its eigenstates with a lesser studied integrable model based on a reflection group $C_{2}$ \cite{gaudin1971_386,footnote1}%
, a symmetry group of a square. As a result, \emph{a spatially odd incident wave of dimers shows a strictly vanishing rate of a monomer production, no matter how far above the dissociation
threshold the incident energy is}.

{ \it Model.} In this work, we investigate the scattering states of a bosonic dimer on a potential barrier by modeling the system as two one-dimensional $\delta$-interacting bosons in presence of a $\delta$-potential barrier located at the center of the system. The Hamiltonian reads: 
\begin{eqnarray}
\begin{split}
\hat{H}=
-\frac{\hbar^2}{2m} \frac{\partial^2}{\partial x_{1}^2}
-\frac{\hbar^2}{2m} \frac{\partial^2}{\partial x_{2}^2} 
+ g_{\text{ab}} \delta(x_{1}) + g_{\text{ab}} \delta(x_{2}) 
\\ +
g \delta(x_{1}-x_{2}) 
\,\,,
\end{split}
\label{H}
\end{eqnarray}
with $g_{\text{ab}} > 0$ being the strength of the potential barrier, ${g=-\hbar^2/a\mu}<0$ the strength of the attractive interatomic interactions and where $a$ is the one 
dimensional even-wave scattering length \cite{olshanii1998_938}, $\mu=m/2$ the reduced mass and $m$ the mass of the atoms. 

Prior to an encounter with the barrier, the energy of the dimer reads
\begin{align}
E=\frac{\hbar^2 k^2}{2m} + \frac{\hbar^2 k^2}{2m} - \frac{\hbar^2}{2\mu a^2}\,\,
\label{energy_condition}
\end{align}
where $\hbar k$ is the momentum of \emph{either} particle in the dimer; the latter is related to the center-of-mass incident momentum, $\hbar K$, as $\hbar k \equiv \hbar K/2$.
The dissociation condition, $k>1/a$, can be readily inferred from the relationship (\ref{energy_condition}): it ensures that the system has enough kinetic energy to 
invest towards dissociation \cite{gaudin1983_book_english}.

The presence of the barrier in \eref{H} breaks the integrability of the model. However, we show below that the spatially odd scattering states of the system described by \eref{H} coincide with the ones of the integrable model described by the following Hamiltonian:
\begin{eqnarray}
\begin{split}
\hat{H}_{C_{2}}=
-\frac{\hbar^2}{2m} \frac{\partial^2}{\partial x_{1}^2}
-\frac{\hbar^2}{2m} \frac{\partial^2}{\partial x_{2}^2}
+ g_{\text{ab}} \delta(x_{1}) + g_{\text{ab}} \delta(x_{2}) 
\\
 +
 g \delta(x_{1}-x_{2}) + g \delta(x_{1}+x_{2}) 
\,\,,  
\end{split}
\label{H_C2}
\end{eqnarray}
The above  Hamiltonian can be shown to be integrable---in all symmetry sectors---using a Bethe Ansatz based on a symmetry group of a square, 
$C_{2}$. This model, as well as all its multidimensional and affine generalizations, was first analyzed by Gaudin \cite{gaudin1971_386,gaudin1983_book_english}, albeit with a conjecture that there exists a single integrability-supporting value of the ratio between the coupling constants $g_{\text{ab}}$ and $g$. The works \cite{emsiz2006_191,emsiz2010_61} show that integrability persists for any ratio between the constants, but restrict the treatment to the ``identity'' representation of the group. The paper \cite{gutkin1982_1} can be used to construct other representations of the corresponding reflection groups, in particular the one we are using in this Letter. In the affine case, it can be shown that the model supports three independent coupling constants \cite{affine}. Note that, the same rich choice of parameters is paralleled in the Calogero-Sutherland-Moser models \cite{perelomov1990_book}.

Let us now show that the spatially odd sector of eigenstates of the Hamiltonians given in Eqs.~(\ref{H}) and (\ref{H_C2}) indeed coincide.
We introduce the permutation,  $\hat{P} \psi(x_{1},\,x_{2}) = \psi(x_{2},\,x_{1})$, and the spatial reflection, $\hat{P'} \psi(x_{1},\,x_{2}) = \psi(-x_{1},\,-x_{2})$, transformations.
They allow us to  introduce a subspace of bosonic, spatially odd states, $\psi_{+,-}$: 
\begin{align}
\hat{P} \psi_{+,-}(x_{1},\,x_{2}) = +\psi_{+,-}(x_{1},\,x_{2})\\
\hat{P'} \psi_{+,-}(x_{1},\,x_{2}) = -\psi_{+,-}(x_{1},\,x_{2}).
\end{align}
Note that  $[\hat{P},\, \hat{H}] = [\hat{P'},\, \hat{H}] = 0$ and $[\hat{P},\, \hat{P'}] = 0$. The reflection symmetry with respect to the $x_2=-x_1$ line---one of the four symmetry axes of the 
$C_2$ model---can be written as $\hat{R}=\hat{P'}\hat{P}$; it commutes with the transformations $\hat{P}$ and $\hat{P'}$ and with the 
Hamiltonian (\ref{H_C2}). As a result, the $\psi_{+,-}$ eigenstates of the Hamilonian are, at the same time, odd eigenstates of the reflection $\hat{R}$, i.e.\
$\hat{R}\psi_{+,-}(x_1,\,x_2)=-\psi_{+,-}(x_1,\,x_2)$. One readily concludes that $\psi_{+,-}(x_1,\,x_2)$ vanishes for $x_1=-x_2$, i.e.\ on the line where the (unphysical) integrability-restoring term $g \delta(x_1+x_2)$ in (\ref{H_C2}) acts. Hence, the spatially odd states $\psi_{+,-}$ are  simultaneously eigenstates of Eqs.~(\ref{H}) and (\ref{H_C2}), allowing us to connect the eigenstates of an integrable system, with all the corresponding conservation quantities associated to them, with the ones of a generally non-integrable one.

Among all the eigenstates of  Eq.~(\ref{H}), we shall now focus on those corresponding to scattering states, i.e.\ non-normalizable states satisfying  plane-wave incoming boundary conditions. 
Since our scattering potential conserves the spatial parity, we will be considering the even and the odd partial waves separately, along with the even and odd scattering solutions. We will show below 
that for the odd states, the hidden 
\emph{partial integrability} revealed above leads to tangible consequences.

The most general form of an even scattering solution of \eref{H} with  respect to the spatial reflection symmetry $\hat{P}'$ reads
\begin{align}
&
 \psi_{\text{even}}(x_{1},\,x_{2})\hspace{-1mm} \stackrel{X\to \pm \infty}{=} 
 \label{even_scattering_solution}
 \\
 &
 \,
 a^{-\frac{1}{2}} e^{-|x|/a}  \left( \cos[K X] + f_{\text{even}}(K) e^{i K |X|}\right)
+ F_{\text{even}}(\varphi) e^{i \kappa r} 
 \nonumber 
\,.  
\end{align}
with center-of-mass coordinate $X \equiv (x_{1}+x_{2})/2$,  relative coordinates $x \equiv x_{1}-x_{2}$ and incoming center-of-mass momentum $\pm K$, and where we use cylindrical coordinates $(r,\varphi)$, with  $x_{1} = r \cos(\varphi)$ and $x_{2} = r \sin(\varphi)$, for the part of the wavefunction corresponding to unbound monomers. The even monomer scattering amplitude satisfies  
$F_{\text{even}}(\pi+\varphi) = F_{\text{even}}(\varphi)$.
Likewise,  the odd scattering solution reads:
\begin{align}
&
\psi_{\text{odd}}(x_{1},\,x_{2})\hspace{-1mm}  \stackrel{X\to \pm\infty}{=}   
a^{-\frac{1}{2}} e^{-|x|/a}  \bigg( \sin[K X] - 
\label{odd_scattering_solution}
\\
&
\qquad
i f_{\text{odd}}(K) \text{sign}(X) e^{i K |X|}\bigg)  - i F_{\text{odd}}(\varphi) e^{i \kappa r} 
\nonumber
\,\,,
\end{align}
where
$F_{\text{odd}}(\pi +\varphi) = -F_{\text{odd}}(\varphi)$. Here and below, $\kappa =\sqrt{2((ka)^2-1)}/a$. Note also that due to the bosonic symmetry of the incident wave, the monomer scattering amplitudes obey $F_{\text{even(odd)}}(\pi/2-\varphi) = F_{\text{even(odd)}}(\varphi)$%
\footnote{In both cases, the normalization is chosen in such a way that number of atoms falling on the barrier per unit time is the same 
as in unidirectional atomic beam of atoms with a velocity ${\cal V} = \hbar K/4m$ and a unit atomic number density.}%
.

{\it Preservation of chemical composition.} 
Bethe Ansatz integrable systems are known to preserve the chemical composition \cite{yurovsky2006_163201}. 
In particular, in the $C_2$-integrable 
model (\ref{H_C2}), any purely dimeric incident wave will not produce unbound monomers after a collision with the barrier, even at 
energies higher than the dimer dissociation threshold
\footnote{Note that the unphysical delocalized dimers centered about the $x_2=-x_1$ axis are not generally disallowed, but they 
are independently forbidden in the spatially odd states.}%
.  This preservation of the chemical composition can be demonstrated
by considering the available 
rapidities produced by a dimeric incident wave, all of which being substantially complex and as such, supporting no monomers.
Indeed, the underlying reflection group, induced by four mirrors with a $45^{\circ}$ angle between them 
can only permute and change sign of the incoming rapidities, but it is not capable of altering their imaginary parts.  
In particular, the above conclusion is valid for a spatially odd linear combination of the incident dimeric waves.
The scattering solution 
induced by it will also be spatially odd. But as we have shown above, 
odd eigenstates of the Hamiltonian (\ref{H_C2}) are, at the same time, the eigenstates of the empirically relevant Hamiltonian (\ref{H}). 
This brings us to the central result of this Letter: $F_{\text{odd}}(\varphi)$ is \emph{identically zero} at all incident energies.

Hence, for both Hamiltonians, the corresponding odd scattering solution can be written as 
\begin{align}
\begin{split}
&
\psi(x_{1},\,x_{2}) \stackrel{X\to \pm\infty}{=} 
\\
&
\qquad
a^{-\frac{1}{2}}  e^{-|x|/a} \text{sign}(X) \sin[K |X| + \delta_{\text{odd}}(K)]
\,\,.
\end{split}
\label{odd_scattering_solution__no_monomers}
\end{align}
The scattering phase $\delta_{\text{odd}}(K)$ can be obtained after a long but  straightforward 
calculation that mirrors  the one for the $A_{2}$ reflection group  \cite{mcguire1964_622} (scattering of a dimer on a monomer for three distinguishable particles of the same mass, interacting 
with the same strength):
%
\begin{widetext}
\begin{eqnarray}
\delta_{\text{odd}}(K) = 
\frac{1}{2}
\text{arctan}\!\left[
\frac{2 a^2 k \left(a \left(a a_{\text{ab}} k^2-1\right)+a_{\text{ab}}\right)}{(a (a k (a_{\text{ab}} k-1)-1)+a_{\text{ab}}) (a (a k
   (a_{\text{ab}} k+1)-1)+a_{\text{ab}})}
                    \right]
\,\,,
\label{delta}
\end{eqnarray}
\end{widetext}
where $a_{\text{ab}} \equiv -\frac{\hbar^2}{m g_{\text{ab}}}$ is the scattering length associated with the interaction of a single particle with the barrier.
This result also allows  to  define the dimer-barrier odd scattering length as
%
\begin{align}
a_{\text{db},\,\text{odd}} 
= -\frac{d}{dK} \delta_{\text{odd}}(K\!=\!0)
=\frac{a^2}{2(a-a_{\text{ab}} )}
\,\,.
\end{align}

%
%
%

{\it Single incident dimeric wavepacket and its dissociation.} By numerical solution of the time-dependent many-body Schr\"{o}dinger equation associated to  \eref{H}, 
we investigate the scattering of an incident (from the left) dimer, of the form 
\begin{equation}
\psi_{\to}\!(x_1,x_2,t\!=\!0)\propto e^{-|x|/a} e^{-(X+x_0)^2/4 \sigma^2+i K X}.
\end{equation}
where $x=-x_0$ is the position of the center of mass of the dimer at initial time. 
%
\begin{figure}
\includegraphics[width=1.0\linewidth]{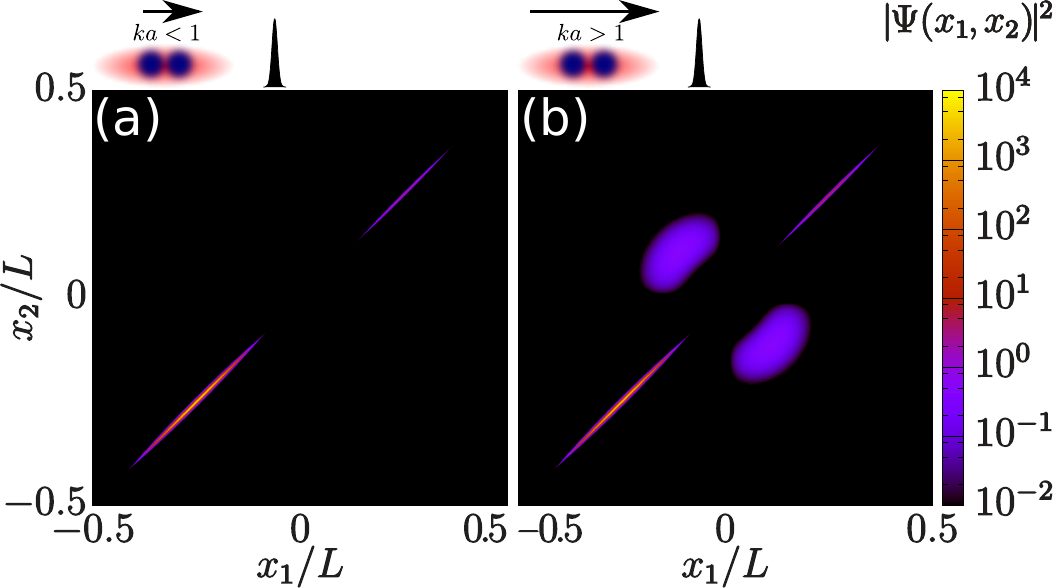}
\caption{Probability density after the collision of a single incident dimeric wavepacket against a barrier, as a function of the dimensionless particle coordinates $x_1/L$ and $x_2/L$, for  incoming momentum  $ka=7/8$ (panel (a)) and  $ka=9/8$ (panel (b)). 
In both panels: the relative strength of the barrier is $g_{ab}/|g|=2$; the dimeric wave packet of an initial width $\sigma/L=1/64$ starts at $-x_0$ with $x_0/L=1/4$ and propagates for $t=mL/(2\hbar k)$,
i.e., the time that would have taken the wavepacket to traveled a distance of $L/2$ if no barrier was present.
On top of each figure we show a schematic representation of the initial state.
}\label{fig:fig2}
\end{figure}

\begin{figure}
\includegraphics[width=0.75\linewidth]{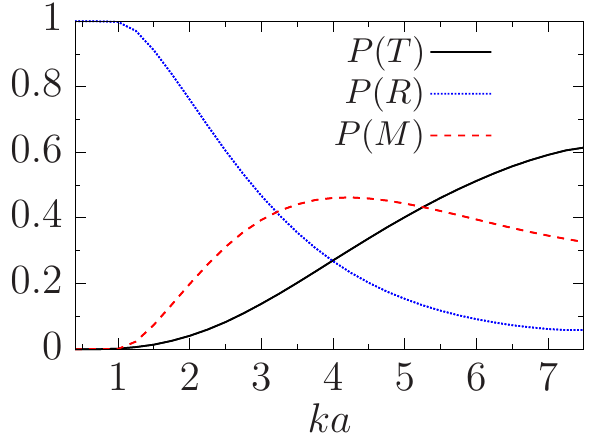}
\caption{Transmission and reflection coefficients of dimers as well as the monomer production coefficient (defined in the main text) for an input unidirectional dimeric wave, as a function of $ka$. 
The other  parameters are the same as in  \fref{fig:fig2}.
}
\label{fig:fig1}
\end{figure}
Figure \ref{fig:fig2} shows the particle density, $|\Psi(x_1,x_2)|^2$, at long times after that the initial dimer has collided against the barrier. If the kinetic energy of the incoming dimer is smaller than the threshold of monomer formation, we observe that the dimer is partially reflected and partially transmitted. However, a ``deflection'' via formation of monomers is clearly visible above threshold.

For each region  $S$ of the $(x_1,x_2)$ plane, the probability  $P(S)=\int_{S} dx_1\,dx_2\,|\Psi(x_1,x_2)|^2$ with $S$ being the four sectors defined as $R \equiv \{x_1<0,x_2<0\}$, 
$T \equiv \{x_1<0,x_2>0\}$, $M \equiv \{x_1>0,x_2<0\} \cup \{x_1>0,x_2>0\}$ yields  the  transmission and reflection coefficients, $P(T)$ and $P(R)$, respectively, along with the monomer formation probability, $P(M)$. Our results for the three coefficients as a function of the initial wavevector of the dimer  are summarized in Fig.~\ref{fig:fig1}. We notice that the non-vanishing monomer production indeed requires $k >1/a$. 

{\it Two counterpropagating dimeric wavepackets.}
We consider next the solution of the two-body Schr\"{o}dinger equation when the following initial condition is taken:
\begin{align*}
&
\psi(x_1,x_2,t\!=\!0)=
\\
&
\quad
e^{-i\phi/2}\psi_{\to}(x_1,x_2,t\!=\!0) + e^{i\phi/2}\psi_{\gets}(x_1,x_2,t\!=\!0)\!\!\!
\end{align*}
where
${\psi_{\gets}\!(x_1,x_2,0)\!\propto\! e^{-|x|/a} e^{-(X-x_0)^2/4 \sigma^2 - i  K X}}$. 

\begin{figure}[t]
\includegraphics[width=1.0\linewidth]{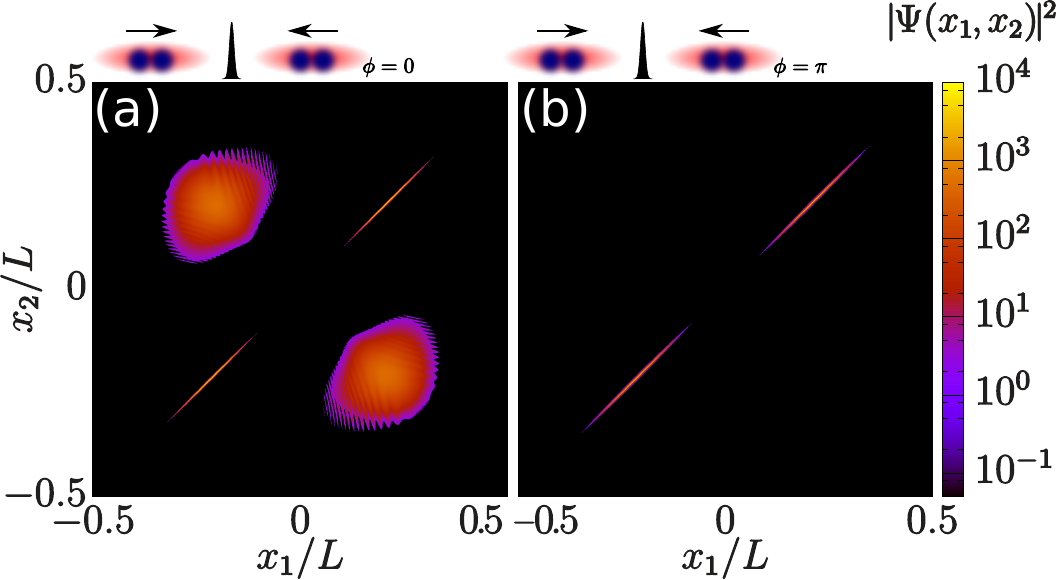}
\caption{Probability density after the collision of the barrier as a function of the dimensionless coordinates  $x_1/L$ and $x_2/L$ for  a linear superposition of two motional states of the dimer with input kinetic energies higher than the dissociation energy ($ka=4$) and an initial relative phase between dimers $\phi=0$ (a) and $\phi=\pi$ (b). In (a) we observe how monomers are produced while in (b) they probability density is completely suppressed. The other parameters are the same as in \fref{fig:fig2}. On top of each figure we show a schematic representation of the initial state.}
\label{fig:fig3}
\end{figure}

\begin{figure}[t]
\includegraphics[width=0.75\linewidth]{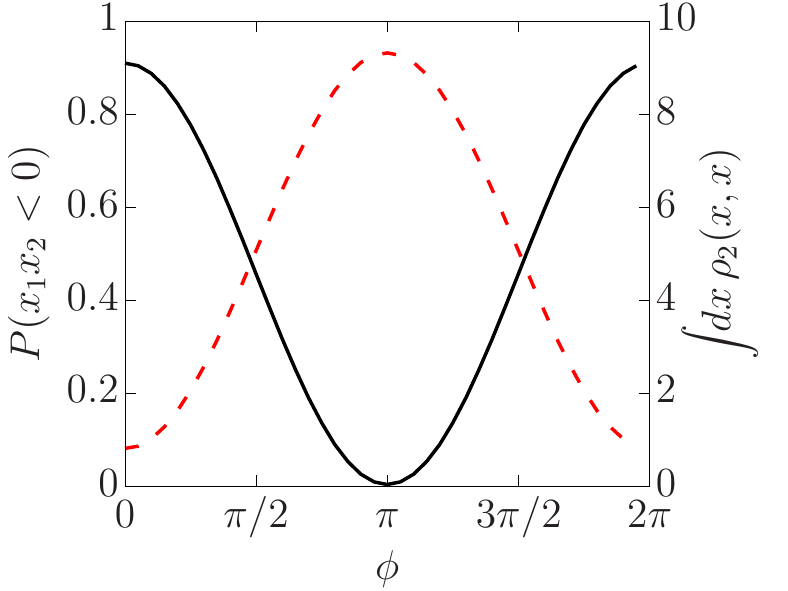}
\caption{Total monomer production after the collision of a linear superposition of two motional states of the dimer against the barrier as a function of the relative phase between the two contrapropagating dimeric wavepackets. The integral shown on the right axis is proportional to the empirically relevant (see \cite{kinoshita2005}) 
two-body correlation function $g_{2}$.
The input kinetic energy is set to two times the dissociation energy ($k a=4$). The other parameters are the same as in  \fref{fig:fig2}.}
\label{fig:fig4}
\end{figure}

In \fref{fig:fig3} we show the probability density when the input state is represented by a spatially even and a spatially odd linear 
combinations of dimeric wavepackets, thus corresponding to the choices $\phi=0$ and $\phi=\pi$ respectively. 
The figure corresponds to the case where  $ka=4$, i.e.\ the input kinetic energy of each dimeric wavepacket is above the threshold for monomer formation.
The figure clearly shows that, while monomers are created in the spatially even configuration, a complete suppression of output monomers is achieved when choosing a spatially odd configuration, in full agreement with the predictions obtained by the spatial odd sector of the integrable $C_2$ model. 
The monomer formation probability as a function or the phase difference $\phi$ between the incident dimeric wavepackets is shown at \fref{fig:fig4}. 
In the same figure we also show a measure proportional to the density-density correlation function $\rho_2(x,y)$, taken at zero distance and  averaged over the sample, ie  $\int dx \rho_2(x,x)$. Notice that
the local second-order correlation $g_2(0)$
has been already made experimentally accessible in a one-dimensional setting \cite{kinoshita2005}.

%


{\it Potential experimental realizations.} 
One-dimensional dimers appear in several areas of physics  of ultracold atoms. One realization is offered by the 
waveguide-trapped spin-$\frac{1}{2}$ fermions  \cite{moritz2005}. 
While the two atoms constituting a dimer are formally distinguishable, the dimer state 
belongs to the bosonic sector of the model,
and so will the scattering state with a moving dimer as the incident wave. In this case, the $\delta$-interaction model is well justified in the 
regime where the size of the ground transverse vibrational state in the guide greatly exceeds the three-dimensional scattering length \cite{olshanii1998_938}.  

Apriori, the one-dimensional dimers described in this Letter can be constructed using any type  of one-dimensional bosonic particles provided that the corresponding interaction potential is sufficiently shallow; quantitatively, it will be required that  the width of the interaction potential $w$ exceeds its scattering length $a$ \cite{flugge1994}. A remarkable example is offered by the recently 
realized dimers of  Rydberg polaritons \cite{liang2017_01478}, where the $w$ is greater than $a$ by at least an order of magnitude.

In both cases considered above, a narrowly focussed sheet of light can be used to generate a fixed one-particle barrier. A similar 
requirement, $w_{\text{ab}} \gg |a_{\text{ab}}|$, must be applied to the sheet waist $w_{\text{ab}} $.

{\it Potential applications.}
One may regard the process of a collision between two dimeric wavepackets and the barrier as a recombination process in an atom interferometer. Indeed, the intensity of all three 
output channels of the scattering event---right moving dimers, let moving dimer, and the monomer production---are expected to depend periodically 
on the relative phase between the input packets (see \fref{fig:fig4} for the latter). 
Unlike the first two, the dimer production is a new possibility. 

Recall that in a chip-based atom accelerometer \cite{wang2005_090405}, the interferometer arms need not be spatially separated: however the readout still requires the separation 
between the channels thus expanding the minimal size of the device. We suggest that in a dimer-based interferometer, the read-out stage of the process can also be made compact if 
the total dimer population, accessible through the two-body correlation function \cite{kinoshita2005} is used as an output.

{\it Summary.} In this Letter we have shown that for short-range-attractive-interacting one-dimensional bosonic atoms, scattering of a spatially odd motional state of a dimer off a barrier produces---even above the dissociation threshold---no unbound atoms. This prohibition originates from a map---valid in the bosonic, spatially odd sector of the Hilbert space---between the 
Hamiltonian of the system and a known, generally unphysical, Bethe Ansatz integrable Hamiltonian associated with the symmetries of a square. Potential experimental realizations include the waveguide 
confined atomic dimers and bound states of two Ridberg polarons. We also suggest that in the context of chip-based atom accelerometers, 
using the monomer production---accessible through the second-order
local correlation function  $g_2(0)$ right after recombination---as an output channel may allow to further miniaturize the readout.

\acknowledgements
We thank Steven G. Jackson and Vladan Vuleti\'{c} for valuable discussions. 
This work is supported by French state funds
ANR-10-LABX-51-01 (Labex LANEF du Programme
d'Investissements  d’Avenir), and ANR-15-CE30-0012-02 (ANR SuperRing project). 
M.O.\ acknowledges financial support provided jointly by the National Science
Foundation, through grant PHY-1607221, and the Binational
(US-Israel) Science Foundation, through grant No. 2015616.


\bibliography{Bethe_ansatz_v024,thermalization_literature_v035,Nonlinear_PDEs_and_SUSY_v031,paper_specific_02,anna,references}

\end{document}